# A new alloy for Al-chalcogen system: AlSe surface alloy on Al (111)

En-Ze Shao[#], Kai Liu[#], Hao Xie, Kaiqi Geng, Keke Bai, Jinglan Qiu, Jing Wang, Wen-Xiao Wang* and Juntao Song*

College of Physics and Hebei Advanced Thin Films Laboratory, Hebei Normal University, Shijiazhuang 050024, China

**ABSTRACT** Metal chalcogenide is a promising material for studying novel underlying physical phenomena and nanoelectronics applications. Here, we systematically investigate the crystal structure and electronic properties of the AlSe surface alloy on Al (111) using scanning tunneling microscopy, angle-resolved photoelectron spectrometer, and first-principle calculations. We reveal that the AlSe surface alloy possesses a hexagonal closed-packed structure. The AlSe surface alloy comprises two atomic sublayers (Se sublayer and Al sublayer) with 1.16 Å along the z direction. The dispersion shows two hole-like bands for AlSe surface alloy located at about -2.2±0.006 eV, far below the Fermi level, which is sharply different from other metal chalcogenide and binary alloys. These two bands mainly derive from the in-plane orbital of AlSe ($p_x$ and $p_y$). These results provide implications for related Al-chalcogen interface. Meanwhile, AlSe alloy have an advantage of large-scale atomic flatness and a wide band gap near the Fermi level in serving as an interface for two-dimensional materials.

**INTRODUCTION**

Transition metal dichalcogenides (TMDCs) combined with transition metals, like Mo or W, and chalcogen elements, such as S, Se, or Te, have emerged as promising materials for diverse applications ranging from optics, nanoelectronics, sensing, and others. TMDCs include several materials with semiconducting and metallic properties. Semiconducting TMDCs like $MoS_2$ could serve as many high quantum efficiency optoelectronic and valleytronic devices [1-3]. The metal TMDCs such as $NbSe_2$ and $TaSe_2$ have attracted immense attention due to rich physical properties (superconductivity, charge density wave (CDW), etc) [1, 4-7].

Other than metal dichalcogenides, metal monochalcogenide alloys are also emerging, such as CuSe, AgTe, and AgSe[8-11]. They exhibit abundant structures and physical properties. CuSe, AgTe and AgSe have remarkable two-dimensional properties with metal atoms and chalcogenide in a plane forming hexagonal structure. And there are also metal and chalcogenide alloys with structures of chains, for example, Te on Cu (111) system with $2\sqrt{3} \times \sqrt{3}$ or $5\sqrt{3} \times \sqrt{3}$ Cu-Te chains [13-16]. For electronic properties, recently, it is reported that CuSe and AgTe possess two dimentional Dirac nodal line fermions (its Dirac point extend along $M - \Gamma - K$ high-symmetry line, forming Dirac nodal line), protected by mirror reflection symmetry[8, 9, 11]. Moreover, the chalcogen-based surface alloys are promising to be used as substrates. For example, CuSe alloy could tune the CDW properties of $TiSe_2$ as a substrate for $TiSe_2$[12]. As a common metal substrate, aluminum is simple and cheap. And the Al-chalcogen compounds are also intermediate products in Al-ion battery. So far, the chalcogen -Al system has not been sufficiently studied. And there are controversy for S/Al (111) and Se/Al (111) alloy among scientists that whether the structure of AlS and AlSe is planar or buckling [17-19]. It is reported monolayer AlSe is an interface between Si substrate and $Al_2Se_3$ [20, 21]. However, identifying the electronic properties of AlSe is a challenge due to the complex energy bands of the Si substrate. Here, we take Se/Al (111) alloy as an example to clarify the formation process and structure of chalcogen elements/Al (111) system.

In this study, we systematically investigate the crystal structure and electronic properties of AlSe alloy on the Al (111) surface combining experiment with calculation. The atomic arrangement of AlSe on Al (111) is directly revealed using high-resolution scanning tunneling microscopy (STM). Reflection high-energy electron diffraction (RHEED) and X-ray photoelectron spectroscopy (XPS) are also carried out to monitor the formation process of AlSe alloy and its chemical bonding. The interplay between the substrate and AlSe alloy is analyzed using angle-resolved photoemission spectroscopy (ARPES) and density functional theory (DFT) calculations. Our results show a buckled structure of AlSe alloy with a band gap near the Fermi level.

## RESULTS AND DISCUSSION

Figure 1 shows the typical XPS spectra of AlSe alloy from the core level of Se 3d. There are two distinct peaks of Se $3d_{3/2}$ and Se $3d_{5/2}$ due to the spin-orbit interactions. In our experiment, the lowest panel of Figure. 1(a) shows the spectrum of Se 3d after deposition at room temperature. The characteristic peaks of Se can be fitted as four spectroscopic contributions shown in Figure 1(a). The two red peaks are labeled as $Se_{b1}$ and $Se_{b2}$ at a binding energy of 56.10 eV and 55.31 eV, corresponding to the levels related to bulk Se atoms, in perfect accord with previous studies[10]. That means they come from chemical bonds characterized by Se-Se in Se bulk. The two blue peaks are labeled as $Se_{a1}$ and $Se_{a2}$ at a binding energy of 55.30 eV and 54.60 eV, whose binding energy is shifted by $\Delta E_{chem} = 0.8\ eV$ to lower binding energy. It indicates a charge transfer process between substrate and Se atoms, after Se atoms deposition on Al (111). In other words, Se-Al bonds were formed during Se adsorbing on Al (111) substrate. So, the two chemical environments for Se deposition on Al (111) at room temperature include Se-Se and Se-Al bonds. The annealing process was taken out to get a single chemical environment. First, samples were annealed at 200 °C, which significantly changed the photoemission spectrum. The intensity of the $Se_{a1}$ and $Se_{a2}$ peaks increased. In contrast, $Se_{b1}$ and $Se_{b2}$ peaks (indicated Se-Se bond) were reduced, suggesting that the annealing process either evaporated the excess Se overlayer or promoted Se interaction with Al . Then the annealing temperature was elevated to 370 °C, the peaks ascribed to the Se-Se bond ($Se_{b1}$ and $Se_{b2}$ peaks) entirely vanished. The only remaining peaks were related to the Se-Al bond ($Se_{a1}$ and $Se_{a2}$ peaks), signifying a single phase of the AlSe alloy layer formed on the surface. The XPS spectra confirmed the surface Al atoms of the substrate reconstructed with Se. Moreover, the uniform XPS spectra taken at different area of samples demonstrate high qulity of AlSe alloy.

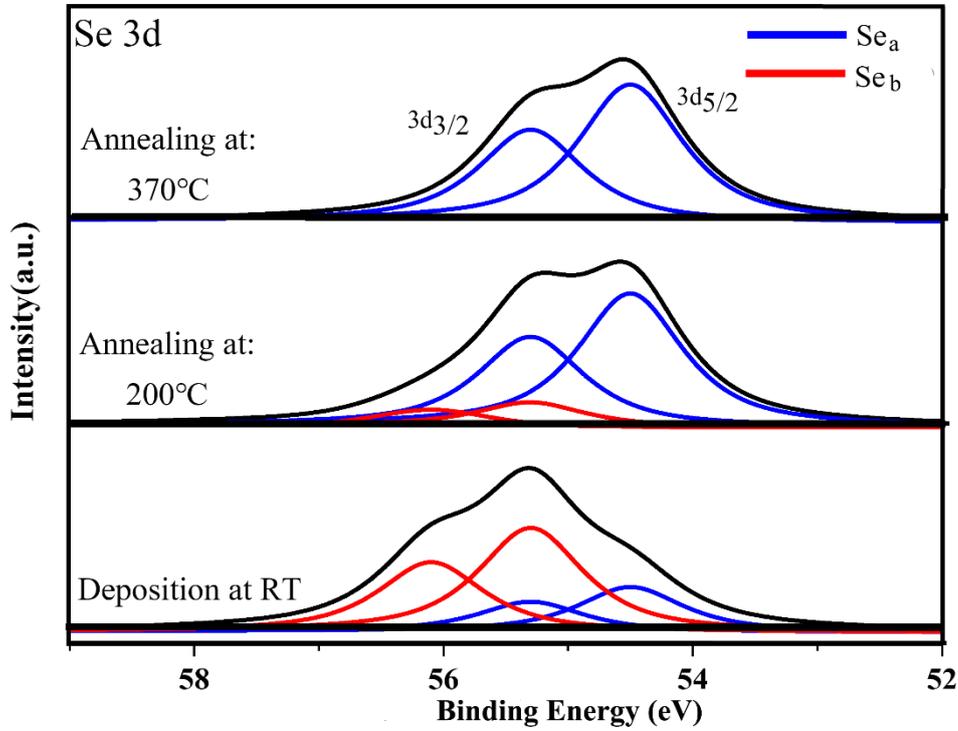

Figure1. XPS spectra for Te 3d and Al 2p core levels at normal emission. The lower spectrum is obtained immediately after deposition. The upper two spectra are taken on the same sample after annealing at 200 °C and 370 °C, respectively. Red and blue solid peaks indicate the Voigt-type fitting results of the different chemical states of Se. The red peak represents Se bulk contribution ($Se_b$), and the blue represents Se of AlSe alloy contribution ($Se_a$).

On the other hand, the crystal symmetry of AlSe alloy was detected by RHEED. Figure. 2 (a) and (b) show typical RHEED pictures before deposition. The stripes are bright and sharp resulting from lattice diffraction along $<1\bar{1}0>$ and $<11\bar{2}>$ of Al (111) surface, which means a clean and smooth surface. The deposition process was monitored by a film thickness monitor. When the thickness is about 0.75 monolayer (ML), additional streaks appear in the RHEED pictures and become sharper after annealing at 370 ℃, indicated by red arrows in Figure. 2 (c, d). These new stripes are ascribed to the additional ordered film of AlSe alloy on the substrate surface. Only one set of patterns suggests the single direction of AlSe domains. During measurement, streaks induced by AlSe alloy and Al (111) substrate reappear when the sample is rotating by every 60º. So, AlSe alloy also has three-fold symmetry. In addition, the space between streaks of AlSe along Al $<11\bar{2}>$ direction, is $\sqrt{3}$ times larger than that along Al $<1\bar{1}0>$ direction. Such a feature is same as that of Al (111) surface with a hexagonal close-packed structure. These results imply the possibility of hexagonal close-packed structure for AlSe surface alloy and the alignment of crystal orientation for AlSe alloy and Al (111) surface in an epitaxial manner. Moreover, the ratio of stripes spacing for Al (111) surface to AlSe alloy film along Al $<11\bar{2}>$ direction is estimated to be 1.337 ± 0.03, which indicates the ratio of lattice spacing for AlSe alloy to Al (0.286 nm) along $<11\bar{2}>$ direction is 1.337 ± 0.03. Accordingly, the lattice spacing of AlSe is preliminarily estimated to be about $0.382 \pm 0.09\ nm$ ($0.286\ nm \times = 0.382 \pm 0.009\ nm$) along Al $<11\bar{2}>$. Similarly, for Al $<1\bar{1}0>$ direction, the lattice spacing of AlSe is deduced as 0.376±0.006 nm. The analysis of RHEED streaks verify smooth AlSe alloy on a large scale aligning with Al (111) surface. We will disscuss the crystallographic structures in more detail in the following by a combination of STM and DFT.

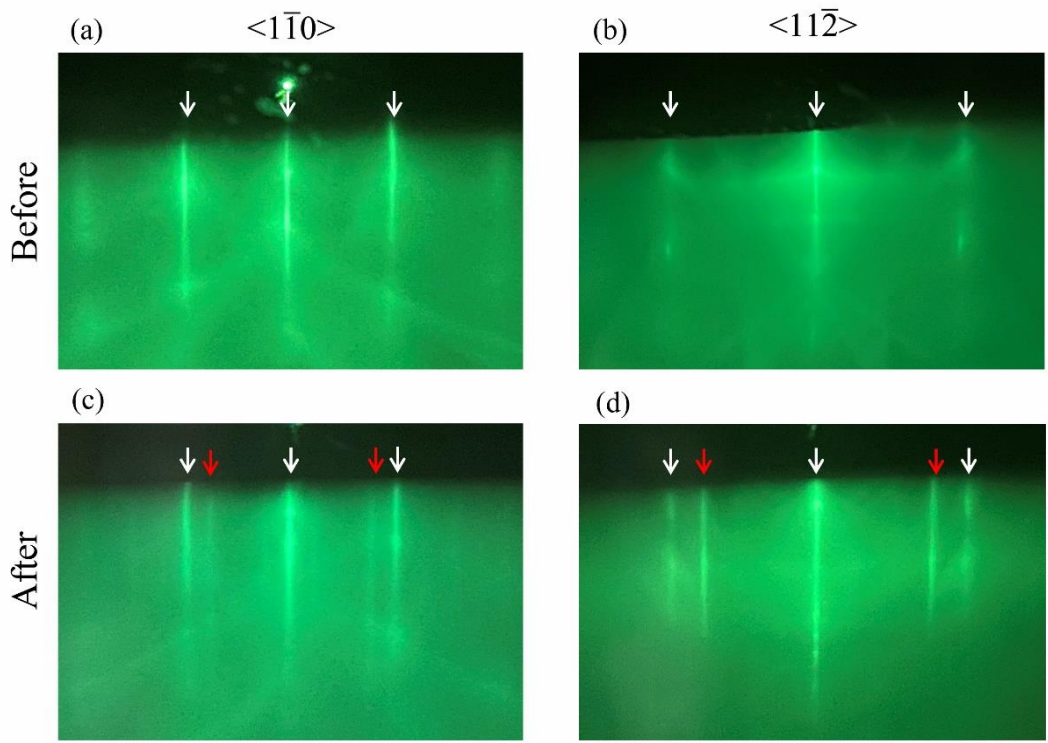

Figure 2. RHEED images observed for Al (111) substrate along (a)<1$\bar{1}$0> azimuth, (b) <11$\bar{2}$> azimuth. White arrows indicate the diffraction pattern from the Al (111) substrate. (c) and (d) RHEED images were observed after deposition and annealing at 370 °C in the same direction as (a) and (b). Red arrows indicate new patterns resulting from AlSe alloy.

STM measurements was taken to further confirm the symmetry and exact atomic structure of AlSe alloy. Figure. 3(a) shows the atoms of AlSe alloy arranged in a hexagonal structure, which is the same as the structure of S atoms on the Al (111) substrate reported previously[19]. The profile in Figure. 3(c) shows a period of 0.38 nm, agreeing well with RHEED results. The AlSe film is distributed continuously across the terrors of the substrate from the STM images. So, all the steps are estimated to be 0.22 nm high, attributed to the height of Al (111), as illustrated in Figures. 3(b, d). However, we did not find the isolated AlSe islands, suggesting they prefer to distribute over a large surface scale. Such a feature has also been observed for Te deposited on Ag (111)[8, 11], and many other alloys[22-24]. These results also provide a proof of the atomically-flat AlSe alloy on Al (111) surfaces, which is one of advantages for AlSe to be used as substrates.

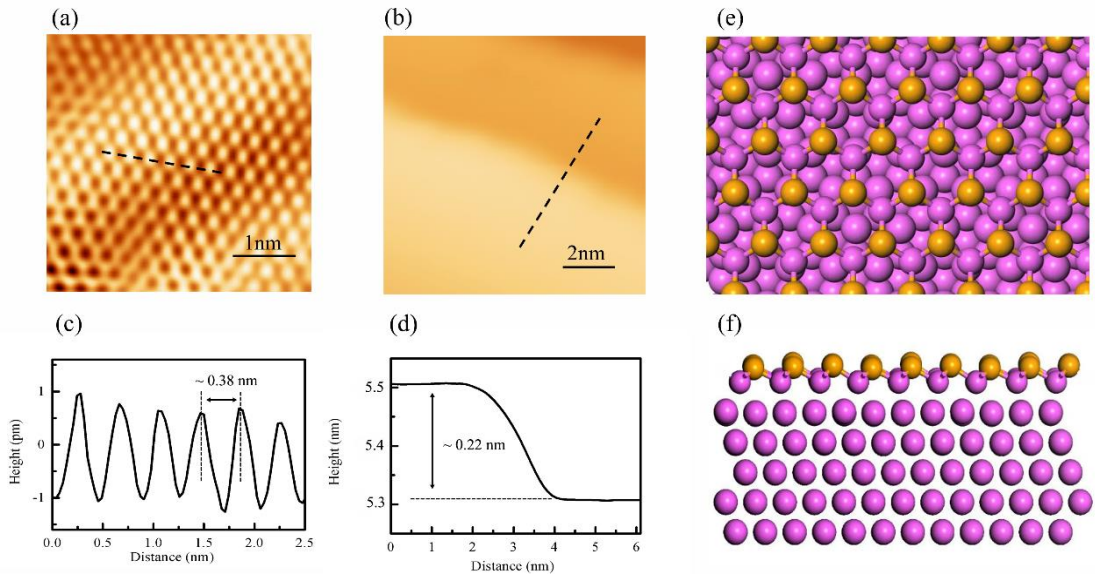

Figure 3. Atomic structure of AlSe surface alloy on Al (111). (a) Typical atomic STM images of AlSe alloy on Al (111) ($V_s$=1.1 V, I=130 pA). (b) STM topology of a terrace of AlSe surface alloy on Al (111). All the surfaces are covered by AlSe alloy ($V_b$=1.1 V, I=130 pA). (c) and (d) Height

profiles from the black dashed lines in (a) and (b), respectively. The length of the nearest atoms is 0.38 nm. (e) and (f) Top and side views of buckled AlSe alloy model.

First-principles calculations further confirm the hexagonal closed-packed structure of the AlSe alloy. A slab model, including the substrate, is shown in Figure. S2 and Figure. S3 in the Supporting Information. The fully relaxed AlSe has two stable phases: the planar phase with a lattice constant of 4.4 Å and buckled phase with a lattice constant of 3.82 Å. We can ascertain the buckled phase of our AlSe sample from two aspects. One, from calculation, buckled phase of AlSe has lower formation energy than the planar phase (see Figure. S2). So the buckled phase is preferential in terms of thermodynamics. Second, the lattice constant of buckled AlSe phase in calculation well agrees with STM and LEED results. The $3 \times 3$ supercell for buckled AlSe (3.82 Å$\times$3=11.46 Å) nearly commensurate with $4 \times 4$ Al (111) surface supercell (2.86 Å$\times$4=11.44 Å). There is only a 0.17% mismatch. These excellent agreement of experiments and calculations confirms the successful synthesis of the buckled AlSe alloy with high quality. The top and side views of the optimized AlSe model are shown in Figures. 3(e) and (f). Different from planar CuTe and AgSe alloy with a $(\sqrt{3} \times \sqrt{3})R30°$ structure, AlSe alloy has a buckled structure including two atomic sublayers, Al and Se layers, with a height of 1.16 Å, as shown in Figure. 3(e)-(f). Based on its unique atomic structure, it is worthwhile to reveal the electronic property.

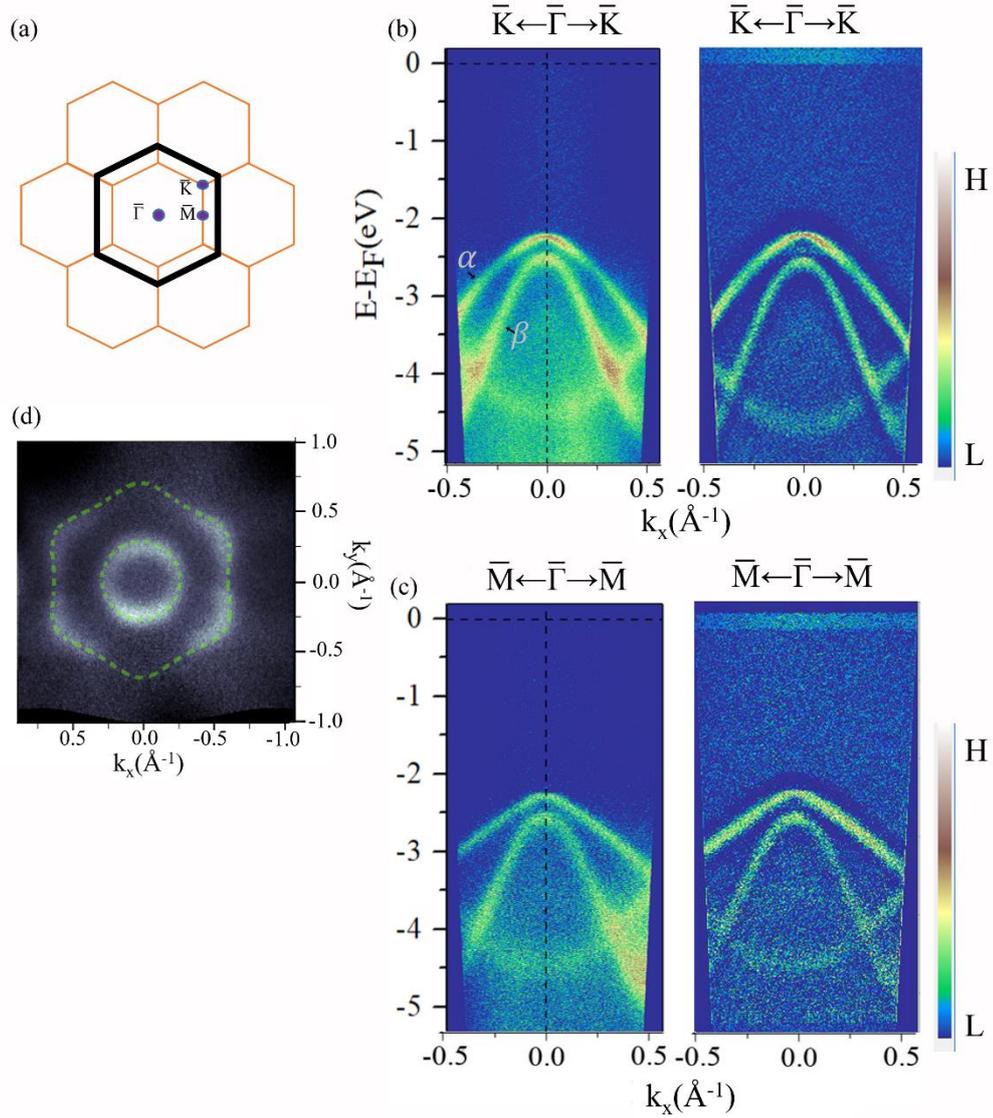

Figure 4. The electronic structure of the AlSe layer on Al (111), probed by ARPES with photon energy of 21.2 eV. (a) The schematic surface Brillouin zone of the Al (111) substrate is shown in a black hexagon. Yellow hexagons represent the SBZs of the AlSe alloy. (b) and (c) left panel: Energy bands mapped along the $\bar{K}-\bar{\Gamma}-\bar{K}$, $\bar{M}-\bar{\Gamma}-\bar{M}$. Right panel: Corresponding maps of the second derivative of the ARPES intensity to the in-plane momentum. (d) The constant energy mapping of AlSe alloy at -3 eV.

Figure. 4 shows the band structure of a single AlSe layer on Al (111), probed by ARPES with photon energy of 21.2 eV. The electron-like band, with its bottom at about -4.6±0.042 eV at Γ point, is the surface-state band of Al (111), consistent with the previous results by ARPES [1]. After being covered by AlSe alloy, the Al (111) substrate still shows a clear surface state band both along $\overline{K\Gamma K}$ and $\overline{M\Gamma M}$ directions, shown in Figure. 4(a-c). The previous literature also reported that the surface state of Al (111) still appeared after being covered by a few layers of films[25-27]. Other than surface state of Al(111) substrate, there are two new hole-like bands labeled α and β along $\overline{K\Gamma K}$ and $\overline{M\Gamma M}$ directions both with a parabolic-type dispersion, which is attribute to the reconstruction of Al (111) surface with Se. The slop of dispersion for both α and β bands along $\overline{K\Gamma K}$ is larger than that $\overline{M\Gamma M}$ directions. The local maximum of the α band is about $2.22 \pm 0.006$ eV below the Fermi level at $\overline{\Gamma}$ point. The energy gap between the maximum of α band and β band is 0.29±0.013 eV. The β band "crossing" the surface state at $-4 \pm 0.011$ eV leads to its strong intensity, which may result from the resonance between the surface state of Al (111) and β band. Furthermore, the ARPES didn't resolve Rashba type spin splitting for AlSe due to the weak Rashiba SOC of the small atomic number of both Al and Se. To better understand the structure of dispersive bands, we mapped the constant energy of bands shown in Figure. 4(c). The constant energy contour was taken at $-3 \pm 0.34$ eV below Fermi energy centered at the $\overline{\Gamma}$ point of the first surface Brillouin zone. The constant energy contour also clearly shows α and β bands. The outer has a flower shape corresponding to the α band. The inner circle is attributed to the β band. The feature of two parabolic-like bands is also typical among other metal chalcogenide and binary alloys. The difference of AlSe from others is that the top of its hole-like bands is very far away from the Fermi level. This feature means the semiconductor character of AlSe alloy on Al (111). If AlSe alloy serves as an interface between metal substrates and their overlayer materials, it may avoid strong interactions introduced by the metal substrates, such as avoiding the hybridization of electronic states near the Fermi level.

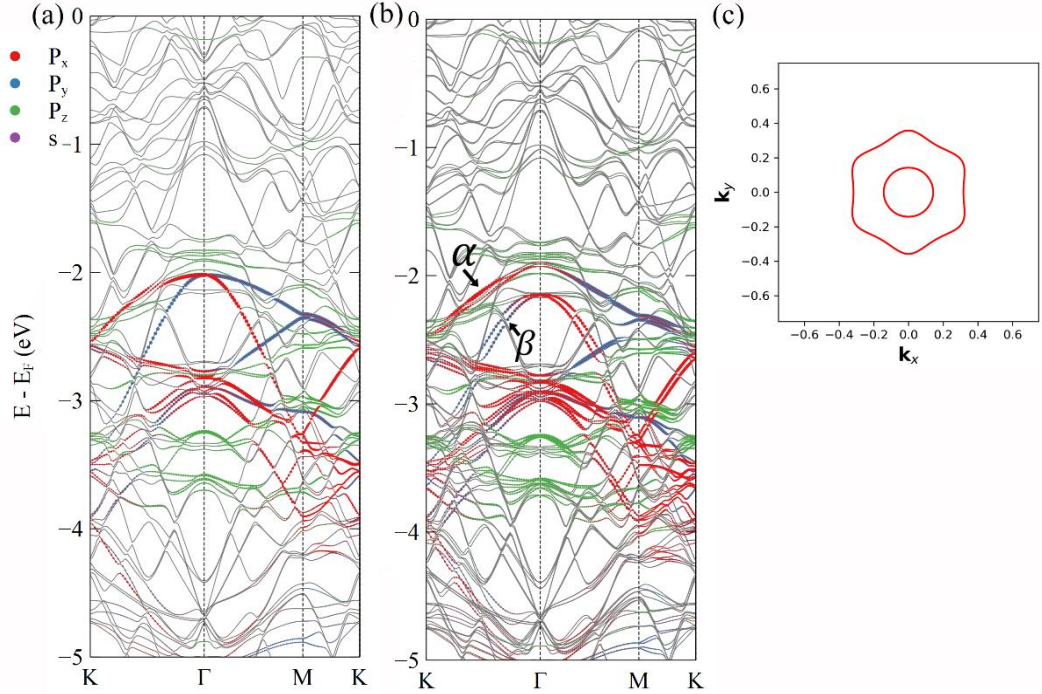

Figure 5. The calculated surface electronic structure of relaxed AlSe alloy on Al (111) (a) without and (b) with SOC. The red, blue, green and purple show projected band structures of top layer AlSe, correspond to the contributions from $p_x$, $p_y$, $p_z$ and s orbital, respectively. The grey is contributed by the substrate. (c) Constant energy contours were calculated for the relaxed AlSe at -2.5 eV.

To investigate the origin of these energy bands, DFT calculations of dispersions for monolayer buckled AlSe alloy on Al (111), along $\overline{K\Gamma MK}$ are shown in Figure. 5. The weights of different orbitals ($p_x$, $p_y$, $p_z$, s) for AlSe alloy are colored in red, blue, green, and purple. The calculated dispersions show that two hole-like bands are contributed by AlSe alloy, which well reproduced the experiment results. Around -3 eV the bands seem like crossing at $\overline{\Gamma}$ point, which results from the back-folding of α and β bands due to the reduced surface Brillouin zone in the calculation (For more discussion see Surpporting Information). The projected band structures (Figure. 5a) show that α /β bands are mainly contributed by the in-plane orbitals of AlSe ($p_x$, $p_y$). Further, with SOC the α and β bands split away 0.24 eV also in accordance with experimental separation of $0.29 \pm 0.013$ eV. What's more, the band structure mapping at -3 eV around Γ point shown in Figure. 5(c) quite well reproduce the experimental result indicated in Figure. 4(d), which is very similar to other binary alloys [10, 23, 24]. By carefully comparing the calculational and experimental bands, we find the absolute energy location of α /β bands in the calculation is 0.3 eV, higher than the experimental data (2.22 ± 0.006 eV), which is comparable to the case of AgTe. This difference possibly attribute to regardless of self-energy effects in the calculation.

For better understanding the bands structure of AlSe, we also calculated the isolated AlSe layer shown in Figure. S4. Other than α and β bands, there is one more narrow band across the Fermi level labeled by black arrows in Figure. S4(a), which is attributed to the out-of-plane orbitals of AlSe ($p_z$ orbital). However, it's absent in experiment. It is because the out-of-plane $p_z$ orbital is sensitive to the coupling of substrate and easily hybridized. Due to the strong coupling of AlSe alloy and substrate, the $p_z$ orbital is hybridized and become indistinguishable from substrate states, demonstrated by comparing the bands without and with substrates shown in Figure. S4 and Figure. 5. Such a feature is the same as other metal chalcogenides on metal substrates reported previously[9-11]. According to the projected bands of isolated AlSe layer, the upward dispertions in the range of the Fermi level to -6 eV are all related to $p_z$ orbital. These bands are hybridized and become indistinguishable from substrate states demonstrated by our calculation and experiment shown in Figure. 4 and Figure. 5. In addition, previous literature reported that the isolated AlSe alloy has an electron-like band labeled $\gamma$ band crossing α and β bands[17, 18]. In contrast, we did not find the electron-like bands of AlSe alloy in the experiment and calculation. There are two reasons. First they didn't consider the effect of substrate which may significantly couple with the overlayer and thus change its electronic properties. Second, they use the planar structure that Al and Se are in the same height. However, from our experimental and calculational results, the buckled structure is more reasonable for AlSe on Al(111).

Recently, it has been reported that CuSe as an interface for $TiSe_2$ could tune the CDW properties[12], demonstrating that the metal chalcogenide is a good candidate as an interface to tune the properties of 2D materials. Because of the atomically-thin structure of 2D materials and their high surface-to-volume ratio, the interfaces between 2D and substrates significantly affect the electronic properties. The substrate, especially metal substrates, may bring charge injection, strain, and charge carrier trapping to 2D material. Atomic flat and dielectric substrates are beneficial for obtaining the intrinsic properties of 2D materials and transport measurements such as hBN substrate. Contrary to its similar metal chalcogenide alloys such as CuSe, AlSe alloy on Al (111) possesses a large band gap and a flat atomic surface on the whole substrate. Based on these characteristics, AlSe alloy may be used as an intermediate transition layer to avoid the strong interaction between the metal substrate and 2D materials and to approach the intrinsic properties of 2D materials, which is a promising direction for future research.

**CONCLUSIONS**

In summary, we utilized high-resolution STM and ARPES to investigate the formation process of selenium on Al (111) in detail and perform a comprehensive study of the structure and electronic properties of AlSe alloy. XPS shows that the binding energy of Se shifts to lower energy to form a single chemistry condition of Al-Se bonds. The Al atoms on the surface reconstruct and bond with Se to form a buckled AlSe alloy. According to STM and RHEED, AlSe has a close hexagonal packing arrangement aligning with the Al (111) surface. The main band of AlSe at $\bar{\Gamma}$ is two parabolic like bands with negative dispersion. Based on the calculation, these two bands mainly derive from in plane orbital of AlSe ($p_x$ and $p_y$). Furthermore, AlSe is atomic flat on a large scale. It has a wide energy band gap near the Fermi level, which make it potential as an interface to avoid the strong interaction introducing by metal substrate for 2D materials and result in intrinsic 2D materials with excellent properties.

**EXPERIMENTAL METHODS**

**Sample Characterization.** ARPES experiments (Scienta Omicron DA30) were performed with He I hv=21.2 eV radiation from a high-brightness monochromatized commercial helium lamp. The base pressure was maintained below $5\times10^{-11}$ torr during ARPES measurements, and the temperature was 10K. On the other hand, samples were transferred by the UHV suitcase chamber for STM measurement at 77K with an electrochemically etched W tip. The base pressure of the STM system (CreaTec) is better than $8 \times 10^{-11}$ mbar.

**Sythesis of AlSe.** The Al (111) surface was prepared by repeated cycles of Ar+ sputtering and annealing at 450℃ in UHV. The surface cleanliness was controlled by *in-situ* high-resolution XPS on the impurity core levels and RHEED on the sharp and consecutive stripe. XPS was carried out by a DA30 electron energy analyzer (Scienta Omicron) and monochromatized AlK$\alpha$ with 1486.7-eV photon energy and 100-eV pass energy. High-purity Se (99.99%, Sigma-Aldrich) was evaporated from a Knudsen cell with a temperature of 115℃, in MBE chamber with a pressure lower than $3 \times 10^{-10} torr$. The substrate was kept at room temperature during 10-min deposition. The process of deposition was monitored by RHEED and thickness monitor.

**First-Principles Calculations.** First-principles calculations were performed using the Vienna *ab initio* simulation package[28], with the Perdew-Burke-Ernzerhof (PBE) exchange and correlation function[29]. The electron wavefunctions were expanded on a plane wave basis with a kinetic energy cutoff of 400 eV, and the Brillouin zone was sampled by Monkhorst-Pack *k*-mesh with a density of 2π×0.02 Å$^{-1}$. The slab model of monolayer AlSe on top of three layers of Al (111) surface was used for simulating the interfacial interaction between them. A 20 Å vacuum layer was applied along the z-direction to avoid interactions between images.

**Author Contributions**

#E. shao and K. Liu contributed equally.

**ACKNOWLEDGEMENTS**

We thank Hongbo Wu and Xiaochuan Ma for useful discussions. This work is support by the National Natural Science Foundation of China (Grant Nos. 21114044 and 11904076), Natural Science Foundation of Hebei (Grant No. A2019205313), the Key Program of Natural Science Foundation of Hebei Province (Grant No. A2021205024) and Science Foundation of Hebei Normal University (Grant No. L2019B10).